\documentclass[conference]{IEEEtran}

\makeatletter

\def\ps@IEEEtitlepagestyle{%
  \def\@oddfoot{\mycopyrightnotice}%
  \def\@evenfoot{}%
}
\def\mycopyrightnotice{%
  {\parbox{\textwidth}{\scriptsize \textbf{Note:} This is the author's accepted manuscript of the paper published as: T. M. Mahmoud and N. Kaabouch, ``Benchmarking Classical and Quantum Machine Learning for Intrusion Detection Across Multiple Datasets,'' in \textit{Proc. 2026 6th International Conference on Electrical, Computer and Energy Technologies (ICECET)}, Rome, Italy, 2026. The final published version is available via IEEE Xplore at \url{https://doi.org/10.1109/ICECET65726.2026.11632866}.\\[2pt] \copyright~2026 IEEE. Personal use of this material is permitted. Permission from IEEE must be obtained for all other uses, in any current or future media, including reprinting/republishing this material for advertising or promotional purposes, creating new collective works, for resale or redistribution to servers or lists, or reuse of any copyrighted component of this work in other works.}}%
  \gdef\mycopyrightnotice{}
}

\IEEEoverridecommandlockouts
\usepackage{cite}
\usepackage{amsmath,amssymb,amsfonts}
\usepackage{algorithmic}
\usepackage{graphicx}
\usepackage{textcomp}
\usepackage{xcolor}
\def\BibTeX{{\rm B\kern-.05em{\sc i\kern-.025em b}\kern-.08em
    T\kern-.1667em\lower.7ex\hbox{E}\kern-.125emX}}

\usepackage{blindtext}
\usepackage{eso-pic}
\usepackage{eso-pic}

\usepackage{url}
\begin{document}

\title{\vspace*{1cm}Benchmarking Classical and Quantum Machine Learning for Intrusion Detection Across Multiple Datasets}

\author{Taha M. Mahmoud \\ Artificial Intelligence Research (AIR) Center \\ University of North Dakota \\ \texttt{taha.mahmoud@und.edu} 
\and Naima Kaabouch \\ Artificial Intelligence Research (AIR) Center \\ University of North Dakota \\ \texttt{naima.kaabouch@und.edu}
}

\maketitle

\begin{abstract}
Intrusion detection is examined through a comparative benchmark of classical and quantum machine learning models on three datasets: NSL-KDD, UNSW-NB15, and MQTTEEB-D2025. The analysis focuses on how preprocessing decisions, feature selection strategies, and quantum encoding schemes shape model performance on data with differing levels of noise and complexity. A single unified pipeline is applied, covering normalization, class-imbalance handling, dimensionality reduction, and two feature selection methods: Random Forest importance and a quantum-aware approach that combines Quantum Kernel Alignment with Mutual Information (QKA-MI). Four classifiers are compared: Support Vector Machine, Random Forest, Quantum Support Vector Machine, and Pegasos Quantum SVM. The classical models perform consistently across all three datasets, whereas the quantum models depend more strongly on feature representation and kernel alignment. Quantum-aware feature selection yields marked gains for the quantum models, especially on cleaner datasets, while heterogeneous traffic remains difficult. Pegasos Quantum SVM offers a good trade-off between accuracy and computational cost, underscoring the role of preprocessing alignment in practical quantum intrusion detection.

\end{abstract}


\noindent\fbox{\parbox{\dimexpr\columnwidth-2\fboxsep-2\fboxrule}{\footnotesize\textbf{Correction note (November 2026).} A re-analysis of this benchmark after publication found data leakage and implementation differences that change several reported results, most strongly on MQTTEEB-D2025 and UNSW-NB15. The correction note at the end of this manuscript lists each issue and its measured effect. The text before the note is the accepted manuscript, unchanged.}}

\section{Introduction}
Intrusion Detection Systems (IDS) are a key line of defence for modern networks facing fast-changing cyber threats. With network traffic growing in both volume and complexity, classical machine learning (ML) models such as Support Vector Machines (SVM) and Random Forests (RF) remain central to anomaly detection. These models nonetheless struggle with strongly nonlinear patterns, high-dimensional feature spaces, and the intricate correlations common in cybersecurity data. Quantum Machine Learning (QML) has emerged as a promising alternative that may offer richer feature representations by exploiting superposition, entanglement, and expressive quantum kernels.

Research on QML for intrusion detection is still in its infancy. Most studies use a single dataset, test one quantum feature map, or apply only limited preprocessing, which makes it hard to judge how quantum models behave under realistic and varied conditions. Classical and quantum models are also seldom compared under the same preprocessing, sampling, and feature selection settings. Consequently, the literature still lacks a systematic, end-to-end assessment of which preprocessing steps matter most for QML, whether quantum models benefit from the same techniques as classical ones, and how the choice of quantum encoding affects detection accuracy and computational cost.

To address these gaps, a comprehensive benchmarking framework is developed that evaluates classical and quantum IDS models across several datasets and preprocessing pipelines. Three widely used datasets are considered (NSL-KDD, UNSW-NB15, and the fully preprocessed MQTTEEB-D2025), spanning different degrees of feature diversity, noise, class imbalance, and preprocessing demands. Each stage of the pipeline is examined in turn: feature scaling, imbalance correction, Principal Component Analysis (PCA) reduction, sampling, and two feature selection strategies, namely Random Forest (RF) importance and the quantum-aware Quantum Kernel Alignment with Mutual Information (QKA-MI).

In contrast to earlier studies, four quantum feature maps (Pauli, ZZ, Z, and Hybrid Pauli--ZZ) and two quantum classifiers (QSVM and PegasosQSVC) are benchmarked under consistent experimental settings, alongside classical SVM and RF baselines that share the same preprocessing pipeline. Varying the sample size, the number of selected features, the quantum encoding, and the model type yields detailed insight into accuracy, F1-score, recall, training time, and memory usage.

The principal contributions are summarized below:
\begin{itemize}
    \item A unified benchmarking framework for comparing classical and quantum IDS models over three datasets and a range of preprocessing configurations.
    \item An in-depth analysis of how data preparation, including scaling, PCA, Synthetic Minority Oversampling Technique (SMOTE) balancing, and sample-size variation, affects QML performance.
    \item The first use of the QKA-MI feature selector in an IDS setting, demonstrating that aligning features with quantum kernel geometry improves quantum model performance.
    \item A thorough evaluation of quantum feature maps and classifiers that identifies the conditions under which quantum models approach or exceed classical baselines.
    \item Practical, evidence-based recommendations for designing robust QML-based IDS across diverse operational scenarios.
\end{itemize}

The rest of the paper is structured as follows. Section~\ref{sec:related} reviews prior work on quantum machine learning for intrusion detection and identifies open research gaps. Section~\ref{sec:methodology} describes the unified benchmarking methodology, covering the datasets, preprocessing steps, feature-selection strategies, and quantum feature-map configurations. Section~\ref{sec:results} presents the experimental results on all datasets, examining how preprocessing, feature dimensionality, quantum encoding, and sample size influence performance and runtime. Section~\ref{sec:discussion} offers an integrated discussion of the findings together with directions for future improvement, and Section~\ref{sec:conclusion} closes with the main insights and their implications for quantum-enhanced intrusion detection.

\section{Related Work} \label{sec:related}

QML for IDS is still at an early stage, and only a few studies have explored it directly. Nicesio et al.~\cite{nicesioQuantumMachineLearning2023} provided one of the first surveys of QML-IDS research, but their review focused mainly on accuracy and did not discuss data preprocessing steps or methodological detail.

Faker and Cagiltay~\cite{fakerQuantumMachineLearning2024} later expanded this work by mapping 21 QML-IDS studies. They organized quantum models such as QSVM, QNN, and QCNN, and described how they connect with classical machine learning frameworks. However, their study did not address dataset suitability, data imbalance, or time-based evaluation methods.

Several experimental papers have also explored quantum models for IDS, though many share similar limitations. Gouveia and Correia~\cite{gouveiaQuantumEnhancedMachineLearning2020} and Payares and Martinez-Santos~\cite{payaresQuantumMachineLearning2021} compared QSVM and QNN models but used small datasets and measured performance only with accuracy. Zhang et al.~\cite{zhangNetworkAttackTraffic2022} tested QNNs on synthetic network data, while Gong et al.~\cite{gongNetworkAttackDetection2022} applied variational quantum networks on the older KDD’99 dataset. Kalinin and Krundyshev~\cite{kalininSecurityIntrusionDetection2023} combined QSVM and QCNN models on proprietary streaming data. Across these studies, reproducibility, preprocessing detail, and clear comparisons with classical baselines were often missing.

A more recent hybrid framework was proposed by Abreu et al.~\cite{abreuQMLIDSQuantumMachine2024}. Their QML-IDS approach combined QSVM, QCNN, and variational circuits and showed potential improvements over classical models on public IDS datasets. However, their experiments were limited to simulators, and their preprocessing and data-split methods were not well explained. Their work showed that quantum models could be useful for IDS, but many design and evaluation questions remain unanswered.

Beyond IDS, other studies have looked at broader or related applications. Anguera et al.~\cite{anguera_leveraging_2025} explored QML for intrusion detection in software-defined networks and compared quantum and classical models using different preprocessing workflows. Bellante et al.~\cite{bellanteEvaluatingPotentialQuantum2025} analyzed the role of quantum learning methods in cybersecurity and discussed conditions under which quantum models may offer benefits as hardware improves. For specialized domains, Chen et al.~\cite{chenQuantumMachineLearning2025} studied UAV swarm intrusion detection using quantum kernels, QNNs, and hybrid quantum models, testing them under similar preprocessing and noise settings.

Overall, current QML-IDS research shows promise but also inconsistency. Many model types have been tested, including QSVM, QNN, QCNN, and hybrid circuits, yet few studies systematically test how preprocessing, dataset properties (such as imbalance or feature size), or time-based data splits affect results. In addition, most evaluations rely only on accuracy and ignore runtime, memory use, or stability. These gaps highlight the need for a clear and reproducible benchmarking framework—like the one in this work—to study how data preprocessing and dataset characteristics influence QML model performance in intrusion detection.

\section{Methodology} \label{sec:methodology}

This section describes the unified benchmarking framework used to evaluate classical and quantum machine learning models for intrusion detection. The methodology was designed to be reproducible, dataset-agnostic, and suitable for analyzing how different preprocessing strategies influence quantum kernels and learning performance.

\begin{figure}
    \centering
    \includegraphics[width=1\linewidth]{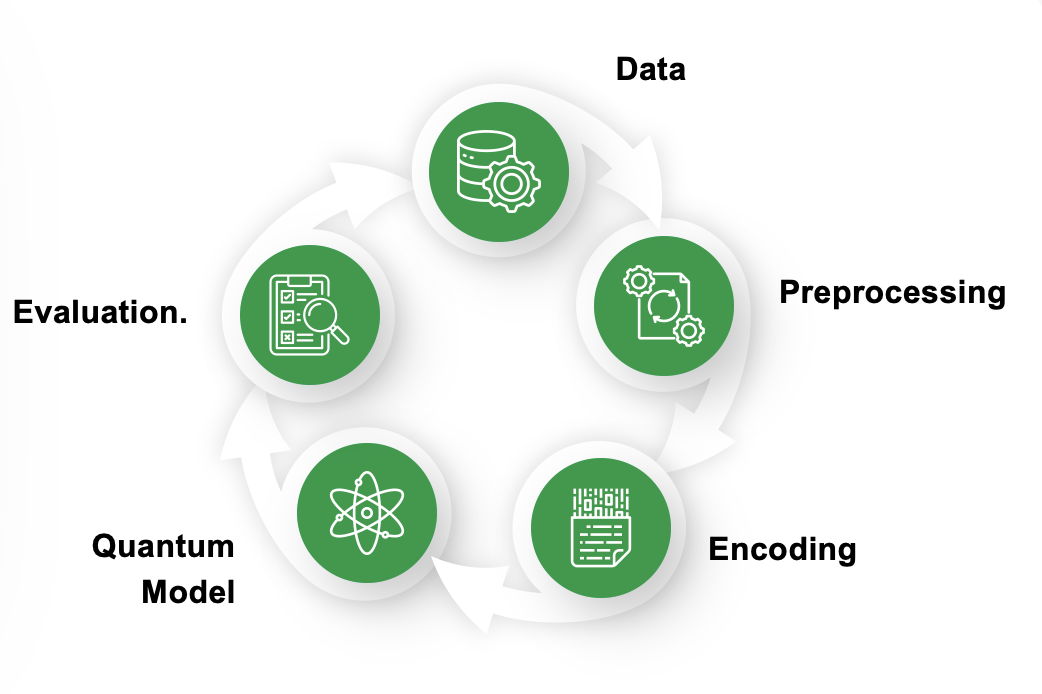}
    \caption{Unified QML Intrusion Detection Pipeline}
    \label{fig:qml_pipe}
\end{figure}

Figure \ref{fig:qml_pipe}   illustrates the unified workflow adopted in this study for benchmarking classical and quantum intrusion detection models. The pipeline begins with the acquisition and preparation of raw network traffic data, which undergoes a series of preprocessing steps such as normalization, PCA reduction, imbalance correction, and feature selection to ensure numerical stability and compatibility with quantum circuits. The processed features are then mapped into a quantum Hilbert space through a chosen encoding strategy, after which the encoded data is used to train either QSVM or PegasosQSVC models. The workflow concludes with a comprehensive evaluation stage, where performance is assessed using accuracy, F1-score, runtime, and memory usage. This end-to-end cycle highlights the iterative and interconnected nature of QML model development, emphasizing that preprocessing and encoding choices directly influence downstream quantum model behavior and overall detection performance.

\subsection{Datasets and Labeling Strategy}

Three publicly available intrusion detection datasets were used in this work: \textit{NSL--KDD}, \textit{UNSW--NB15}, and \textit{MQTTEEB--D2025}. 
NSL--KDD represents traditional TCP/IP traffic and contains moderate feature overlap; 
UNSW--NB15 provides a richer and more heterogeneous view of modern network activity; 
and MQTTEEB--D2025 consists of IoT MQTT telemetry that has already undergone extensive preprocessing.

For UNSW--NB15, the training and testing files were merged into a single dataset before preprocessing.  
MQTTEEB--D2025 includes six attack categories, while the other datasets contain multiple attack types with imbalanced distributions.  
A global configuration parameter, \texttt{LABEL\_MODE}, controls whether experiments are conducted in binary mode (Normal vs.\ Attack) or in multiclass mode.  
Even when training in binary mode, sampling is performed using the original multiclass labels to ensure that rare attack types remain represented in smaller subsets, preserving diversity without inflating computational cost.

\subsection{Data Preprocessing Pipeline}

A unified preprocessing pipeline was applied to all datasets to ensure consistent comparison.  
Categorical attributes, such as protocol and service type, were converted to numerical format using one-hot encoding, while non-informative identifiers (for example, timestamps or connection IDs) were removed to avoid leaking sequential structure into the learning process.  
Instances containing missing values were dropped to ensure that scaling transformations and kernel estimations remained numerically stable.

To study model behavior under constrained data availability—a typical setting in quantum machine learning—stratified sampling was applied with sampling ratios ranging from 0.2\% to 10\%.  
Stratification was always based on the original multiclass labels to prevent the exclusion of minority attack types.  
In datasets exhibiting substantial imbalance, the SMOTE was applied after sampling but before scaling so that synthetic samples correctly reflected the distributional properties of the sampled data.

Feature scaling followed two different pathways depending on the model family.  
Classical models used standard z-score normalization to ensure consistent feature variance.  
Quantum models employed a two-stage pipeline: z-score standardization followed by min–max scaling into the range $[0, 2\pi]$.  
This “angle normalization’’ guarantees that the transformed values correspond to valid rotation angles in parameterized quantum circuits, thereby preventing over-rotation, phase wrapping, and kernel concentration issues.  
A single-stage (direct min–max) alternative was also explored but resulted in less stable quantum kernels.  
Feature dimensionality was optionally reduced using PCA to three components, a choice made to align the classical feature space with three-qubit quantum feature maps while retaining more than 95\% of the total variance across all datasets.

The three-feature (three-qubit) configuration is adopted as a controlled benchmarking setting aligned with current kernel-based quantum circuit widths. This does not imply that intrusion detection is inherently low-dimensional. Classical models can exploit higher-dimensional representations more easily; however, increasing feature dimensionality in quantum models directly increases circuit depth and kernel complexity. Therefore, this constraint enables systematic comparison under a shared qubit-limited regime while preserving over 95\% of data variance.

Feature selection also played a central role in the pipeline.  
Two approaches were examined.  
The first relied on Random Forest (RF) feature importance scores, using the top $k \in \{3, 5\}$ features.  
The second method, Quantum Kernel Alignment with Mutual Information (QKA\_MI), selected features that maximized alignment between the quantum kernel matrix and the class labels. 
The kernel alignment score between a quantum kernel matrix $K$ and label kernel matrix $Y$ is defined as:

\begin{equation}
\mathcal{A}(K, Y) = \frac{\langle K, Y \rangle_F}{\|K\|_F \|Y\|_F}
\end{equation}

where $\langle \cdot , \cdot \rangle_F$ denotes the Frobenius inner product. QKA-MI selects features that maximize this alignment while incorporating mutual information with class labels.

To make QKA\_MI scalable, a time-bounded version incorporating pre-filtering ($M{=}20$ candidates) and kernel subsampling ($N{=}150$ samples) was implemented.

\subsection{Quantum Feature Encoding}

Quantum feature maps were implemented using \textit{Qiskit Machine Learning}.  
The experiments evaluated a set of encoding strategies with varying expressivity, including the PauliFeatureMap, ZZFeatureMap, ZFeatureMap, and a hybrid Pauli–ZZ design that combined rotation and entanglement layers.  
All feature maps used a single repetition (\texttt{reps}=1) to control circuit depth, and three qubits were used throughout to match the PCA-reduced feature dimensionality.  

All quantum computations were executed using the Qiskit \texttt{Sampler} backend on Google Colab, representing a noise-free NISQ-level simulation environment.

\subsection{Model Training}

The benchmarking included two classical models—Support Vector Machine (SVM) with an RBF kernel and a Random Forest classifier with 200 trees—and two quantum models: a QSVM based on the Fidelity Quantum Kernel, and PegasosQSVC, which uses the same quantum kernel but employs stochastic optimization for improved efficiency.  
All models were trained using binary labels $\{0,1\}$ without any relabeling.  
Train–test splits were performed using an 80/20 stratified partition to preserve class proportions across all experiments.

\subsection{Evaluation Metrics}

\begin{figure}[h]
    \centering
    \includegraphics[width=1\linewidth]{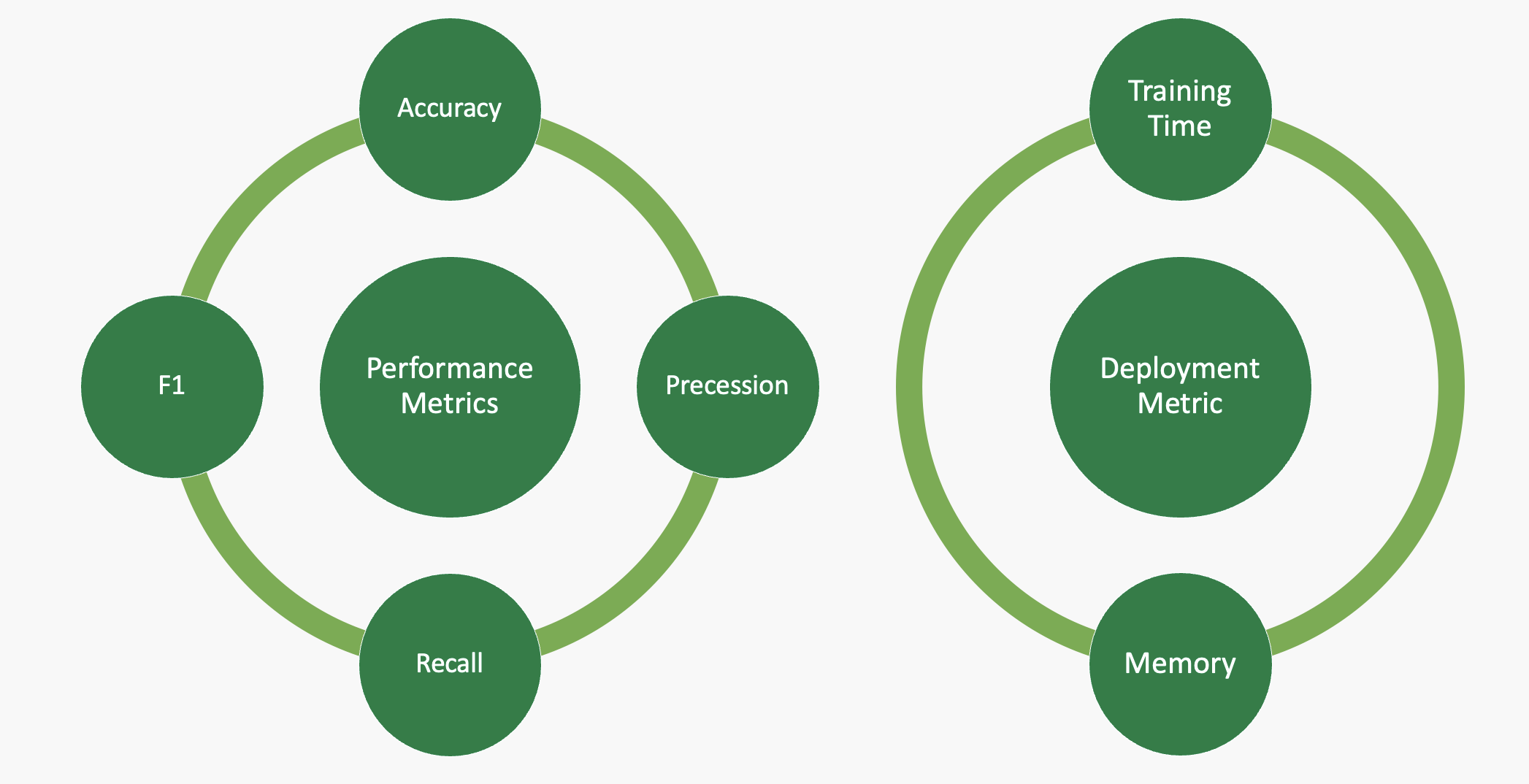}
    \caption{Key performance and deployment metrics used to evaluate classical and quantum IDS models.}
    \label{fig:IDS_EV}
\end{figure}

Performance was assessed using standard classification metrics, including Accuracy, Recall, and F1-score, defined in Equations~\ref{eq:accuracy}--\ref{eq:f1}. To capture computational efficiency, training time and memory consumption were also recorded as shown in Figure~\ref{fig:IDS_EV}. Each experimental configuration---defined by the dataset, feature selector, sampling ratio, number of features, encoding choice, and model---was executed independently to prevent cross-contamination and ensure reproducibility.

To strengthen empirical validity, experiments can be repeated using multiple random seeds and data splits, reporting mean $\pm$ standard deviation for performance metrics. While kernel-based methods executed on deterministic simulator backends exhibit limited stochastic variability, future work will incorporate repeated-run statistical analysis and significance testing (e.g., Wilcoxon signed-rank test) to further validate performance differences.

\begin{equation}
\text{Accuracy} = \frac{TP + TN}{TP + TN + FP + FN}
\label{eq:accuracy}
\end{equation}

\begin{equation}
\text{Precision} = \frac{TP}{TP + FP}
\label{eq:precision}
\end{equation}

\begin{equation}
\text{Recall} = \frac{TP}{TP + FN}
\label{eq:recall}
\end{equation}

\begin{equation}
\text{F1-score} = 2 \times \frac{\text{Precision} \cdot \text{Recall}}{\text{Precision} + \text{Recall}}
\label{eq:f1}
\end{equation}

\subsection{Implementation Environment}

All experiments were conducted in Google Colab (Python 3.12) using Qiskit Machine Learning, scikit-learn, imbalanced-learn, and psutil. All quantum experiments are conducted on a noise-free simulator backend. While this enables controlled comparison of preprocessing and kernel behavior, it does not capture hardware noise, connectivity constraints, or shot-based variability present in current NISQ devices. Deeper feature maps and higher qubit counts are expected to be more sensitive to noise. Evaluating robustness under realistic noise models and hardware execution remains an important extension of this work.
The decision to standardize all pipelines to three numerical features was made to align with the quantum circuit width and does not reflect a limitation of quantum hardware.

\section{Results} \label{sec:results}

This section presents a comprehensive evaluation of classical (SVM, RF) and quantum (QSVM, PegasosQSVC) models across three intrusion detection datasets: NSL--KDD, UNSW--NB15, and MQTTEEB--D2025. All models were trained under identical preprocessing pipelines, including binary labeling, feature standardization, and angle normalization for quantum inputs. The analysis examines the effects of feature selection, feature dimensionality, feature maps, sample size, imbalance handling, and runtime. Results are organized to highlight cross-dataset behavior and the practical trade-offs between accuracy and computational cost.

\subsection{Dataset-Level Performance Comparison}

Table~\ref{tab:overall_results} summarizes the best achieved performance per model and dataset. Classical models remain highly stable across datasets, while quantum models exhibit dataset-dependent variability. PegasosQSVC consistently provides the best accuracy--runtime balance among quantum approaches, whereas QSVM achieves strong accuracy with substantially higher cost.

\begin{table*}[t]
\centering
\caption{Best Achieved Performance of Classical and Quantum Models Across Datasets}
\label{tab:overall_results}
\scriptsize
\begin{tabular}{llcccccc}
\hline
\textbf{Dataset} & \textbf{Model} & \textbf{FS Method} & \textbf{Encoding} &
\textbf{Accuracy} & \textbf{F1-score} & \textbf{Training Time (s)} & \textbf{Observations} \\
\hline
NSL--KDD & SVM/RF & RF/QKA & Classical & 0.90--0.94 & 0.89--0.94 & $<0.2$ & Stable baselines \\
NSL--KDD & PegasosQSVC & QKA\_MI & Pauli & \textbf{0.94} & \textbf{0.94} & 2.9 & Strongest quantum result \\
NSL--KDD & QSVM & QKA\_MI & Pauli & 0.92 & 0.92 & 40 & Accurate but costly \\
\hline
UNSW--NB15 & SVM/RF & RF & Classical & \textbf{0.996} & \textbf{0.995} & $<0.6$ & Classical-dominant dataset \\
UNSW--NB15 & PegasosQSVC & RF & Pauli/Z & 0.96--0.99 & 0.95--0.99 & 1--2 & Competitive \\
UNSW--NB15 & QSVM & RF & Pauli/Z & 0.99 & 0.99 & 1100--1600 & Accurate but slow \\
\hline
MQTTEEB--D2025 & SVM/RF & RF & Classical & \textbf{1.00} & \textbf{1.00} & $<0.3$ & Highly separable dataset \\
MQTTEEB--D2025 & PegasosQSVC & RF/QKA & Pauli/Z & 0.96--1.00 & 0.96--1.00 & 1--7 & Fast, stable \\
MQTTEEB--D2025 & QSVM & RF/QKA & Pauli/Z & 0.98--1.00 & 0.99--1.00 & 400--9000 & Perfect accuracy, heavy cost \\
\hline
\end{tabular}
\end{table*}

\subsection{Effect of Feature Selection and Dimensionality (NSL--KDD)}

Feature selection has the strongest influence on quantum models. Table~\ref{tab:nsl_fs} shows that QKA\_MI substantially improves quantum separability compared to RF selection. Increasing feature dimensionality from $k{=}3$ to $k{=}5$ yields marginal changes for classical models but introduces deeper quantum circuits, sometimes reducing stability.

\begin{table}[t]
\centering
\caption{NSL--KDD: Effect of Feature Selection and Feature Dimensionality}
\label{tab:nsl_fs}
\scriptsize
\begin{tabular}{lcccc}
\hline
\textbf{Model} & \textbf{FS Method} & \textbf{$k$} & \textbf{F1-score} & \textbf{Processing Time (s)} \\
\hline
SVM & RF & 3--5 & 0.90--0.90 & $<0.01$ \\
RF  & RF & 3--5 & 0.94--0.93 & 0.18--0.20 \\
Pegasos & RF & 3--5 & 0.81--0.82 & 4.7--5.2 \\
QSVM & RF & 3--5 & 0.82--0.81 & 73--74 \\
\hline
Pegasos & QKA & 3 & \textbf{0.94} & 2.9 \\
Pegasos & QKA & 5 & 0.86 & 3.1 \\
QSVM & QKA & 3--5 & 0.92--0.86 & 40--89 \\
\hline
\end{tabular}
\end{table}

\textbf{Key finding:} QKA-MI improves quantum models significantly on NSL--KDD (up to +12\% F1), confirming that quantum kernel alignment is effective for datasets with moderate feature overlap.

\subsection{Dataset-Specific and Cross-Dataset Quantum Behavior}

Across the three datasets, quantum and classical models exhibit consistent and interpretable behavior that reflects the underlying data characteristics. NSL--KDD contains moderate noise and overlapping feature distributions, making it a meaningful benchmark for evaluating quantum-aware preprocessing. In this setting, QKA\_MI substantially improves quantum separability, enabling PegasosQSVC and QSVM to achieve competitive performance relative to classical baselines.

In contrast, UNSW--NB15 includes richer, more heterogeneous traffic patterns with higher variance across features. Classical models achieve near-perfect results (F1 $>$ 0.995), while quantum models show larger fluctuations—even under QKA-based feature selection. This pattern aligns with findings in prior QML studies, where quantum kernels are sensitive to heterogeneous feature distributions and may become unstable when data lacks consistent geometric structure.

The MQTTEEB--D2025 dataset behaves differently: it is fully preprocessed, balanced, and highly separable. All models—classical and quantum—achieve near-perfect accuracy and F1-score. Under these conditions, the primary differentiator becomes computational efficiency rather than predictive performance. PegasosQSVC remains efficient and stable, whereas QSVM incurs substantial runtime overhead due to full kernel matrix estimation.

Taken together, three cross-dataset patterns emerge:

\begin{enumerate}
    \item \textbf{Quantum models benefit most from QKA-based feature selection on datasets with overlapping distributions (NSL--KDD).}  
    Well-conditioned, alignment-based features mitigate kernel concentration and improve quantum decision boundaries.

    \item \textbf{Quantum kernels are less stable on heterogeneous, high-variance datasets (UNSW--NB15).}  
    This matches reports in recent literature showing that quantum models are sensitive to feature scaling, variance, and mixed distributions.

    \item \textbf{On clean, preprocessed, highly separable data (MQTTEEB--D2025), both classical and quantum models converge to perfect classification.}  
    In such cases, runtime—not accuracy—is the primary factor determining practicality.
\end{enumerate}

Overall, the results demonstrate that quantum model performance is strongly shaped by dataset characteristics and preprocessing quality. QKA-MI and angle normalization are essential for stabilizing performance on complex datasets but do not provide additional advantage when the dataset is inherently linearly separable.

\subsection{Accuracy--Runtime Trade-off}

\begin{figure}[h]
\centering
\includegraphics[width=1\linewidth]{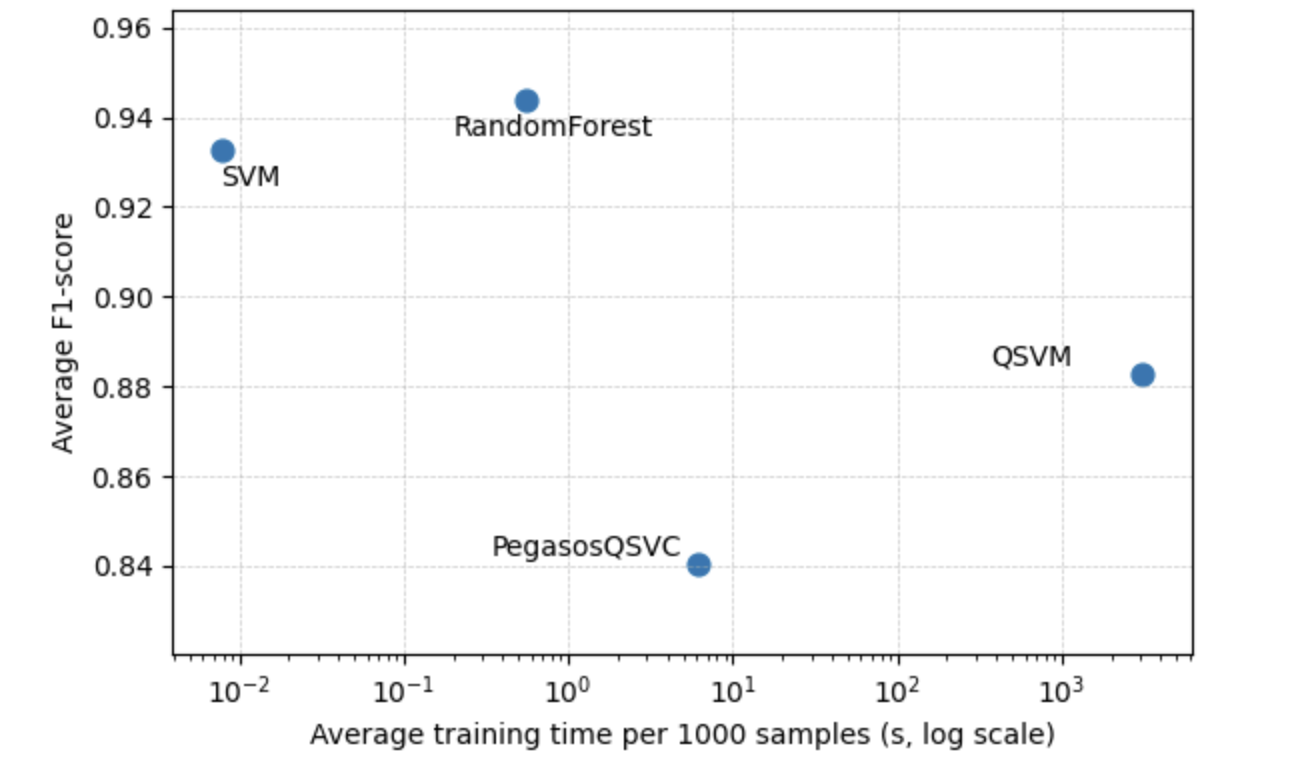}
\caption{Accuracy vs.\ Training Time (log-scale) across aggregated dataset results. PegasosQSVC offers the best accuracy--runtime compromise, while QSVM yields high accuracy but large computational overhead.}
\label{fig:acc_runtime}
\end{figure}

Fig.~\ref{fig:acc_runtime} presents accuracy vs.\ training time on a log scale. Classical models achieve high accuracy with minimal cost. PegasosQSVC stands out as the most practical quantum model. QSVM reaches strong accuracy but with exponential increases in runtime due to full kernel matrix construction, a behavior documented widely in QSVM literature.

\subsection{Impact of Scaling and Imbalance Handling}

Angle normalization to $[0,2\pi]$ consistently stabilizes quantum feature maps. Experiments without angle scaling showed large F1 fluctuations—confirming prior studies that quantum rotations are highly sensitive to unbounded or poorly scaled inputs. SMOTE improves minority recall without harming quantum kernel structure, provided that synthetic samples remain near decision boundaries.

\section{Discussion} \label{sec:discussion}

The experimental results indicate that quantum models can approach classical baselines when preprocessing is carefully aligned with quantum kernel structure. Across all datasets, classical models (SVM and RF) consistently achieved stable and high accuracy with low computational cost. In contrast, quantum models were more sensitive to preprocessing decisions, particularly feature selection, scaling, and dimensionality reduction. This behavior reflects the strong dependence of quantum kernels on the geometry of the encoded feature space.

Although this study focuses on binary classification for controlled benchmarking, the evaluated datasets contain multiple attack categories. Extending quantum kernel methods to full multiclass intrusion detection using one-vs-rest or hierarchical strategies represents an important research direction and may provide further insight into class separability in quantum feature space.

A central finding is the impact of the QKA\_MI feature selector. On NSL--KDD, QKA\_MI improved quantum F1-scores by more than 12\% compared with Random Forest selection. This improvement arises because QKA\_MI ranks features based on alignment with the quantum kernel, promoting separability in Hilbert space. In contrast, RF-based selection prioritizes classical statistical relevance, which does not always translate into improved quantum kernel conditioning.

Feature scaling also proved critical. Mapping inputs to $[0,2\pi]$ stabilized parameterized rotations and reduced kernel variance. Without angle normalization, certain entangling feature maps exhibited inconsistent behavior due to over-rotation and phase distortion. These results highlight that quantum preprocessing must account for both classical variance structure and qubit rotation constraints.

Increasing feature dimensionality from three to five had minimal impact on classical models but reduced stability for quantum models due to deeper circuits and more complex kernel matrices. Scaling to larger feature spaces will require increased qubit counts, which may introduce kernel concentration effects, noise sensitivity, and computational overhead. Future work should explore qubit-efficient encodings and hybrid dimensionality reduction strategies to address these limitations.

Overall, the observed performance differences do not suggest inherent limitations of quantum learning, but rather emphasize the importance of preprocessing alignment and kernel conditioning. Under controlled and preprocessing-aligned configurations, quantum models demonstrated competitive performance, while classical models remained strong and efficient baselines.

\section{Conclusion} \label{sec:conclusion}

This study presented a unified benchmarking framework for evaluating classical and quantum machine learning models for intrusion detection across the NSL–KDD, UNSW–NB15, and MQTTEEB–D2025 datasets. The results confirm that classical models provide stable and efficient baselines, while quantum models can achieve competitive performance when supported by quantum-aware preprocessing, particularly appropriate feature selection, scaling, and controlled dimensionality. Among the evaluated quantum approaches, PegasosQSVC offers the most practical balance between accuracy and computational cost, whereas QSVM incurs substantial runtime overhead. Overall, the findings indicate that current limitations of quantum machine learning in intrusion detection are driven primarily by preprocessing and kernel conditioning challenges rather than inherent model weaknesses, highlighting the importance of preprocessing alignment for the practical adoption of quantum-enhanced intrusion detection systems.

\section*{Correction Note (November 2026)}
After publication, the benchmark pipeline was re-executed under its pinned environment and audited against the CRISP-QCML-IDS review gates. The issues below affect the results as reported. Corrected values are means over 10 seeds with leak-free, fit-on-training-data preprocessing; a corrected and extended benchmark is reported in a follow-up study by the same authors (in preparation).

\begin{enumerate}
\item \textbf{MQTTEEB-D2025 label leakage.} A derived binary label column (\texttt{Attack\_Binary}) remained among the input features and was ranked first by feature importance. Without it, F1-score falls by 0.08 (QSVM), 0.16 (PegasosQSVC), 0.04 (Random Forest) and 0.06 (SVM); the best leak-free values are 0.971 (Random Forest) and 0.950 (QSVM). The statements that all models reach perfect classification on MQTTEEB-D2025 do not hold.
\item \textbf{UNSW-NB15 row index.} The sequential \texttt{id} column shipped with the dataset remained a feature. Because the files are ordered by class, this column alone gives F1 $=0.83$. Removing it lowers Random Forest by 0.014 and raises SVM by 0.032 and QSVM by 0.023; on leak-free data QSVM ties the best classical model (0.907). The near-perfect classical scores on UNSW-NB15 ($F1>0.995$) partly reflect this column.
\item \textbf{Encoding range.} Features were scaled to $[0,2\pi]$ and Qiskit's Z, ZZ and Pauli feature maps apply $P(2x)$, so rotations span $4\pi$ and values $\pi$ apart map to the same state. The scaling therefore causes, rather than prevents, phase wrapping. A bounded range $[0,\pi/2]$ with clipping raises quantum F1-score on NSL-KDD and MQTTEEB-D2025.
\item \textbf{Preprocessing order and split.} Feature selection, SMOTE, scaling and PCA were fitted before the train--test split, and the split was not stratified in the code, contrary to the text. Fitting after the split changes F1-score by up to 0.12 on the most imbalanced dataset.
\item \textbf{Circuit depth and feature maps.} The feature maps were built with two repetitions, not one. The Pauli, ZZ and hybrid maps compile to the same circuit, so two distinct maps (Z and ZZ) were compared, not four.
\item \textbf{QKA-MI comparison.} The 180\,s time budget stopped QKA-MI selection after two features, so the QKA-MI results used two qubits while the Random Forest results used three. The reported gain of up to 12 points is not a like-for-like comparison; with matched settings the gain is 0.05--0.08 F1 under the bounded range and about 0.015 under $[0,2\pi]$.
\item \textbf{PegasosQSVC.} PegasosQSVC was run unseeded, and repeated runs of one configuration range from 0.65 to 0.92 F1. Over 10 seeds it is below QSVM on every dataset, so it is not the strongest quantum model.
\item \textbf{Runtime.} The reported training times measure simulator overhead: the same $200\times200$ kernel takes 40--152\,s through the Qiskit Sampler and under 1\,s from statevectors. Kernel entries and shots are the appropriate cost measure.
\end{enumerate}
NSL-KDD results are least affected: correcting the leakage changed no model by more than 0.008 F1-score.

\bibliographystyle{IEEEtran}
\bibliography{references}

\end{document}